\documentclass[12pt,a4paper]{conference}

\usepackage{fancyhdr}
\usepackage{graphicx,amsmath,amssymb,cite}
\usepackage{multind}
\newcommand{\beq}{\begin{equation}}
\newcommand{\eeq}[1]{\label{#1}\end{equation}}
\newcommand{\eeqn}{\end{equation}}

\newcommand{\beqa}{\begin{eqnarray}}
\newcommand{\eeqa}[1]{\label{#1}\end{eqnarray}}
\newcommand{\eeqan}{\end{eqnarray}}

\let\bar=\overbar

\newcommand{\Dslash}{\not{\hbox{\kern-4pt $D$}}}
\newcommand{\dslash}{\not{\hbox{\kern-2pt $\del$}}}

\newcommand{\msb}{{\bar{\ssstyle M \kern -1pt S}}}

\begin{document}

\Chapter{THE GOTTFRIED SUM RULE IN AN UNQUENCHED QUARK MODEL 
}
           {Gottfried sum rule}
           {R. Bijker \it{et al.}}

\addcontentsline{toc}{chapter}{{\it R. Bijker}} \label{authorStart}

\begin{raggedright}


R. Bijker$^{\star,}$\footnote{Proceedings MENU 2007, eConf C070910, 210}~, 
E. Santopinto$^{\%}$
\bigskip\bigskip


$^{\star}$ICN-UNAM, AP 70-543, 04510 Mexico DF, Mexico\\
$^{\%}$INFN and University of Genova, via Dodecaneso 33, 16164 Genova, Italy\\

\end{raggedright}


\begin{center}
\textbf{Abstract}
\end{center}
We present an unquenched quark model \index{subject}{unquenched quark model} 
for baryons in which the effects of quark-antiquark pairs 
\index{subject}{quark-antiquark pairs} ($u \bar{u}$, $d \bar{d}$ and $s \bar{s}$) 
are taken into account in an explicit form. The method is illustrated with an 
application to the flavor asymmetry \index{subject}{flavor asymmetry} of the 
nucleon sea. 

\index{subject}{Gottfried sum rule}
 
\section{Introduction}

In the constituent quark model (CQM), the proton is described in terms 
of a $uud$ three-quark configuration. A direct handle on higher Fock 
components (such as $uud-q \bar{q}$ configurations) is provided by 
parity-violating electron scattering (PVES) experiments, which have shown 
evidence for a nonvanishing strange quark contribution, albeit small, to 
the charge and magnetization distributions of the proton \cite{Acha}. 
The contribution of strange quarks to the nucleon is of special interest 
because it is exclusively part of the quark-antiquark sea $q \bar{q}=s \bar{s}$. 
Additional evidence for higher Fock components in the proton wave function 
comes from measurements of the $\bar{d}/\bar{u}$ asymmetry in the nucleon 
sea \cite{sea} and from CQM studies of baryon spectroscopy. 
   
Theoretically, the role of higher Fock components in the CQM has been studied 
in \cite{riska}, while the importance of mesonic contributions to the spin 
and flavor structure of the nucleon is reviewed in \cite{Speth}. In another, 
CQM based, approach the effects of $s \bar{s}$ pairs in the proton were  
included in a flux-tube breaking model \cite{baryons}. 

The aim of the present contribution is to discuss the flavor asymmetry of the 
nucleon sea in an unquenched quark model in which the effects of quark-antiquark 
pairs are included in a general and systematic way.   

\section{Unquenched Quark Model}

In the flux-tube model for hadrons, the quark potential model arises from an 
adiabatic approximation to the gluonic degrees of freedom embodied in a flux 
tube \cite{flux}. The impact of quark-antiquark pairs in meson spectroscopy has 
been studied in a flux-tube breaking model \cite{mesons} in which 
the $q \bar{q}$ pair is created with the $^{3}P_0$ quantum numbers of the vacuum.  
Subsequently, it was shown \cite{OZI} that a {\it miraculous}  
set of cancellations between apparently uncorrelated sets of intermediate states 
occurs in such a way that they compensate each other and do not destroy the good 
CQM results for the mesons. In particular, the OZI hierarchy is preserved and 
there is a near immunity of the long-range confining potential, since the change 
in the linear potential due to the creation of quark-antiquark pairs in the string 
can be reabsorbed into a new strength of the linear potential, {\em i.e.} in a new 
string tension. As a result, the net effect of the mass shifts due to pair creation 
is much smaller than the naive expectation of the order of the strong decay widths.  
However, it is necessary to sum over large towers of intermediate 
states to see that the spectrum of the mesons, after unquenching and renormalizing, 
is only weakly perturbed. An important conclusion is that no simple truncation of 
the set of meson loops is able to reproduce such results \cite{OZI}.

The extension of the flux-tube breaking model to baryons requires a proper 
treatment of the permutation symmetry between identical quarks. As a first step, 
Geiger and Isgur investigated the importance of $s \bar{s}$ loops in the proton 
in an unquenched quark model based on an adiabatic treatment of the flux-tube 
dynamics to which the $s \bar{s}$ pair creation with vacuum quantum numbers is 
added as a perturbation \cite{baryons}. 
In the conclusions, the authors emphasized: {\em It also seems very 
worthwhile to extend this calculation to $u \bar{u}$ and $d \bar{d}$ loops. Such an 
extension could reveal the origin of the observed violations of the Gottfried sum 
rule and also complete our understanding of the origin of the spin crisis}.  
In this contribution, we take up the challenge and present a generalization  
of the formalism of \cite{baryons} which now makes it possible to study the effects 
of $q \bar{q}$ pairs in an unquenched quark model (i) for any initial baryon 
(ground state or resonance), (ii) for any flavor of the quark-antiquark pair, and 
(iii) for any model of baryons and mesons, as long as their wave functions are 
expressed in the basis of the harmonic oscillator. 

These extensions were made possible by two developments: the solution of the 
problem of the permutation symmetry between identical quarks by means of 
group-theoretical techniques, and the construction of an algorithm to 
generate a complete set of intermediate states for any model of baryons 
and mesons. 

\section{Flavor Asymmetry}
 
The first clear evidence for the flavor asymmetry of the nucleon sea was provided 
by NMC at CERN \cite{nmc}. The flavor asymmetry is related to the Gottfried integral 
for the difference of the proton and neutron electromagnetic structure functions 
\begin{eqnarray}
S_G = \int_0^1 dx \frac{F_2^p(x)-F_2^n(x)}{x} 
= \frac{1}{3} - \frac{2}{3} \int_0^1 dx \left[ \bar{d}(x) - \bar{u}(x) \right] ~.
\end{eqnarray}
Under the assumption of a flavor symmetric sea, one obtains the 
Gottfried sum rule $S_G=1/3$. The final NMC value is $0.2281 \pm 0.0065$ at $Q^2 = 4$ 
(GeV/c)$^2$ for the Gottfried integral over the range $0.004 \leq x \leq 0.8$ \cite{nmc}, 
which implies a flavor asymmetric sea. The violation of the Gottfried sum rule has been 
confirmed by other experimental collaborations \cite{hermes,nusea}. 
Theoretically, it was shown \cite{Thomas}, that the coupling of the nucleon to the pion 
cloud provides a natural mechanism to produce a flavor asymmetry. 

In the present model, the flavor asymmetry can be 
calculated from the difference of the number of $d$ and $u$ sea quarks in the proton 
\begin{eqnarray}
N_{\bar{d}}-N_{\bar{u}} 
= \int_0^1 dx \left[ \bar{d}(x) - \bar{u}(x) \right] ~. 
\label{asym}
\end{eqnarray}
Note that, even in absence of explicit information on the (anti)quark distribution 
functions, the integrated value can be obtained directly from the left-hand side 
of Eq.~(\ref{asym}). The corresponding value for the Gottfried integral is $0.185$, 
in qualitative agreement with the NMC result. 
It is important to note that in this calculation the parameters were taken from 
the literature \cite{baryons,CR}, and that no attempt was made to optimize their 
values. Since the dependence of the Gottfried integral on the value of these  
parameters and/or different models of baryons and mesons has not yet been 
investigated in detail, the numerical value quoted here is to be regarded 
as preliminary. 

\section{Summary, conclusions and outlook}

We discussed an unquenched quark model for baryons which includes, in addition 
to $s \bar{s}$ loops, the contributions of $u \bar{u}$ and $d \bar{d}$ loops. 
In an application to the flavor asymmetry of the nucleon sea, it was shown 
that the $q \bar{q}$ pairs immediately lead to an excess of $\bar{d}$ 
over $\bar{u}$ quarks in the proton, in agreement with the experimental data. 

In our opinion, the result for the flavor asymmetry is very promising and 
encouraging. We believe that the inclusion of the effects of quark-antiquark 
pairs in a general and consistent way, as suggested in \cite{BS} and in this 
contribution, may provide a major improvement to the constituent quark model, 
increasing considerably its range of applicability. 

\section*{Acknowledgments}

This work was supported in part by a grant from CONACYT, Mexico 
and in part by INFN, Italy.

\end{document}